\documentclass[twocolumn,secnumarabic,amssymb, nobibnotes, aps, prd]{revtex4}
\usepackage{graphicx}

\begin{document}

\narrowtext

{\bf Comment on ``Coherent Detection of Electron Dephasing''}

 \bigskip

The authors of the paper \cite{Strambini} conclude that an Aharonov-Bohm ring with asymmetric electron injection can act as a coherent detector of electron dephasing. This conclusion is based on a theoretical result that electrons can be reflected because of magnetic flux. But this result can not be correct. The reflection because of magnetic flux contradicts to the fundamental law of momentum conservation. The contradiction with one of the fundamental laws of physics in such scandalous form is absent in the Aharonov - Bohm effect \cite{AB1959}, although there is a problem with non-local force free momentum transfer \cite{QCh,Nature08,Book1989}, which has provoked controversy \cite{Nature08,Book1989,Boyer00,Boyer02,PRL07}. In spite of the change in the interference pattern no {\it overall} deflection of electrons is observed in the Aharonov-Bohm effect because of magnetic flux \cite{QCh}. As opposed to this agreement with the conservation law the theoretical result shown on Fig.2 of \cite{Strambini} predicts overall reflection of electrons because of magnetic flux.  

This mistake is a consequence of a false realistic interpretation of the orthodox quantum mechanics. In spite of reiterated notices of Heisenberg {\it that the concept of the probability function does not allow a description of what happens between observations} \cite{Heisenberg1959} the authors \cite{Strambini} use this function for description of electron transmission. The interference pattern observed in the two-slit interference experiment \cite{QCh} can be described with help of the probability function $\Psi = A e^{i(pr + Et)/\hbar }$ of momentum eigenstate. But this function can not describe the transmission of particles through two slits because the probability $P = |\Psi |^{2} = A^{2}$ does not change in time $t$ and space $r$. We can know about the transmission only observing an arrival of electron \cite{QCh} at the detecting screen. When the electron has arrived at a time $t$ we can conclude that it has transmitted through two slits at the time $t - L/v$, where $v = p/m$ is the electron velocity. It is important to note that the electron arrives at a single point $x$ of the detecting screen \cite{QCh} whereas the probability function predicts the arrival with the probability 
$$P(x) = A_{1}^{2}(x) + A_{2}^{2}(x) + 2A_{1}(x)A_{2}(x) \cos(\varphi _{1} - \varphi _{2}) \eqno{(1)} $$ 
at different points. Here $A_{1}(x)$, $A_{2}(x)$ are the amplitude of an arrival probability at the point $x$ of a particle passing through the first, second slit; $\varphi _{1} - \varphi _{2} = \oint_{l} dl \nabla \varphi  = \oint_{l} dl p/\hbar  = \oint_{l} dl (mv+qA)/\hbar = \delta \varphi _{0}(x) + 2\pi \Phi /\Phi _{0}$ is the phase difference along the two paths between a source and a point $x$ of the detecting screen; $2\pi \Phi /\Phi _{0}$ is the phase shift because of a magnetic flux $\Phi = \oint_{l} A dl$; $\Phi _{0} = 2\pi \hbar /q$ is the flux quantum. The orthodox quantum mechanics evades this discrepancy with help of the collapse of the probability function $\Psi $ or {\it 'quantum jump' of our knowledge} \cite{Heisenberg1959} at observation. 

In connection with the mistake made in \cite{Strambini} it is important to note that a magnetic flux $\Phi $ changes the arrival probability at a point $x$ (1) because of the change of the phase difference $\varphi _{1} - \varphi _{2} = \delta \varphi _{0}(x) + 2\pi \Phi /\Phi _{0}$, but not transmission probability $P_{tr} = \int dx P(x) = \int dx (A_{1}^{2} + A_{2}^{2}) = 1$. The transmission probability through the ring arms \cite{Strambini} can not be described with help of the probability function because of an other reason. The phase difference $\varphi _{1} - \varphi _{2} = \oint_{l} dl \nabla \varphi $ can have any value $\delta \varphi _{0}(x) + 2\pi \Phi /\Phi _{0}$ only if the probability function $\Psi = |\Psi | e^{i\varphi }$ collapses at observation. But such observation is not possible in the case of the Aharonov - Bohm ring considered by the authors \cite{Strambini}. The phase difference must be divisible by $2\pi $, 
$\varphi _{1} - \varphi _{2} = \oint_{l} dl \nabla \varphi = 2\pi n $,  because of the requirement that probability function must be single-valued $\Psi = |\Psi | e^{i\varphi } = |\Psi | e^{i(\varphi + 2\pi n )}$ at any point without its collapse. Because of this requirementn the Aharonov - Bohm effect requires the periodical dependencies in magnetic flux $\Phi $ of the persistent current and other values observed in normal metal \cite{Science09} and superconductor \cite{JETPh07} rings. 

 \bigskip

A.V. Nikulov

Institute of Microelectronics Technology 

and High Purity Materials, 

Russian Academy of Sciences, 

142432 Chernogolovka, Moscow District, 

RUSSIA.

\end{document}